\documentclass[aps,prl,secnumarabic,nobibnotes,twocolumn,superscriptaddress,longbibliography]{revtex4-1}

\usepackage{amsfonts}
\usepackage{mathrsfs}
\usepackage{amsmath}
\usepackage{color}
\usepackage{graphicx}
\usepackage{bm}
\usepackage{amssymb}
\usepackage{xspace}
\usepackage{epstopdf}
\usepackage{dcolumn}
\usepackage{longtable}
\usepackage{multirow}
\usepackage[colorlinks=true, letterpaper=true, pdfstartview=FitV, linkcolor=blue, citecolor=blue, urlcolor=blue]{hyperref}

\providecommand{\tabularnewline}{\\}

\makeatother

\begin{document}

\title{Valley-Layer Coupling: A New Design Principle for Valleytronics}
\author{Zhi-Ming Yu}
\thanks{Z.-M. Yu and S. Guan contributed equally to this work}
\address{Research Laboratory for Quantum Materials, Singapore University of
Technology and Design, Singapore 487372, Singapore}

\author{Shan Guan}
\thanks{Z.-M. Yu and S. Guan contributed equally to this work}
\address{Research Laboratory for Quantum Materials, Singapore University of
Technology and Design, Singapore 487372, Singapore}
\affiliation{State Key Laboratory of Superlattices and Microstructures, Institute of Semiconductors, Chinese Academy of Sciences, Beijing 100083, China}

\author{Xian-Lei Sheng}
\address{Department of Physics, Key Laboratory of Micro-nano Measurement-Manipulation and Physics (Ministry of Education), Beihang University, Beijing 100191, China}
\address{Research Laboratory for Quantum Materials, Singapore University of Technology and Design, Singapore 487372, Singapore}

\author{Weibo Gao}
\address{Division of Physics and Applied Physics, School of Physical and Mathematical Sciences, Nanyang Technological University, Singapore 637371, Singapore}
\address{The Photonics Institute and Centre for Disruptive Photonic Technologies, Nanyang Technological University, Singapore 637371, Singapore}

\author{Shengyuan A. Yang}
\address{Research Laboratory for Quantum Materials, Singapore University of
Technology and Design, Singapore 487372, Singapore}


\begin{abstract}
We introduce the concept of valley-layer coupling (VLC) in two-dimensional materials, where the low-energy electronic states in the emergent valleys have valley-contrasted layer polarization such that each state is spatially localized on the top or bottom super-layer. The VLC enables
a direct coupling between valley and gate electric field, opening a new route towards electrically controlled valleytronics.
We analyze the symmetry requirements for the system to host VLC, demonstrate our idea via first-principles calculations and
model analysis of a concrete 2D material example, and show that an electric, continuous, wide-range, and switchable control of valley polarization can be achieved by VLC.
Furthermore, we find that systems with VLC can exhibit other interesting physics, such as valley-contrasting linear dichroism and optical selection of the electric polarization of interlayer excitons.
\end{abstract}
\maketitle

\textit{\textcolor{blue}{Introduction.}} In certain semiconducting crystals, the electronic conduction and/or valence band may feature multiple
energy extremal points in the momentum space, which endows the low-energy carriers with an additional
valley degree of freedom. Analogous to spin, this valley freedom can be used to encode and to process information, giving rise to the concept of valleytronics~\cite{Rycerz2007,GunawanPRL2006,XiaoPRL2007,YaoPRB2008,XiaoPRL2012,ZhuNP2012,CaiPRB2013}. In the past decade, with the discovery of two-dimensional (2D) valleytronic materials such as graphene and transitional metal dichalcogenide monolayers, the field of valleytronics has undergone rapid development and attracted tremendous interest~\cite{xu2014spin,Liu2015CSR,Schaibley2016,Vitale2018}. A variety of valleytronic devices, like valley filters~\cite{Rycerz2007,GunlyckePRL2011,PanPRB2015,PanPRB2015b,JiangPRL2013,Gruji2014,Ju2015,Sui2015gate,Tong2016NC,Nguyen2016}, valley valves~\cite{Akhmerov2008,CrestiPRB2008,San-Jose2009,Qiao2011NL,QiaoPRL2014,Li1149}, and valley splitters~\cite{Garcia-Pomar2008,Settnes2016,Cheng2018}, have been proposed.

For practical valleytronics applications, a crucial prerequisite is the existence of an efficient and controllable method to generate valley polarization, i.e., the valley population imbalance of carriers. So far, in experiment, valley polarization is either generated dynamically, e.g., by optical pumping with circularly polarized light~\cite{mak2012,zeng2012,Cao2012NC,Hsu2015,mak2018light}, by spin injection~\cite{ye2016electrical}, and by valley Hall effect~\cite{Mak1489,shimazaki2015,lee2016}, or induced by an applied magnetic field~\cite{CaiPRB2013,LiPRL2014,aivazian2015,srivastava2015,MacNeillPRL2015,QiPRB2015,jiang2017}. Generally speaking, for device applications, static means are preferred than dynamic means, and fully electric control is preferred than magnetic or optical control. Particularly, static control by gate electric field is most desirable, because of its advantages in compactness, power efficiency, and compatibility with current semiconductor technology. However, such a gate field control of valley polarization has not been realized, which poses an outstanding challenge for the field of valleytronics.

The key reason behind the challenge is symmetry. The valleytronics research has been focused on systems with a pair of valleys connected by the time reversal operation $\mathcal{T}$, such as in graphene and MoS$_2$ monolayer. Hence, to create valley polarization, one has to break the $\mathcal{T}$ symmetry, by a dynamical process or by magnetic field. In contrast, the gate electric field does not break $\mathcal{T}$, which forbids its capability to produce polarization for the $\mathcal{T}$-connected valleys.

\begin{figure}
\includegraphics[width=8.5cm]{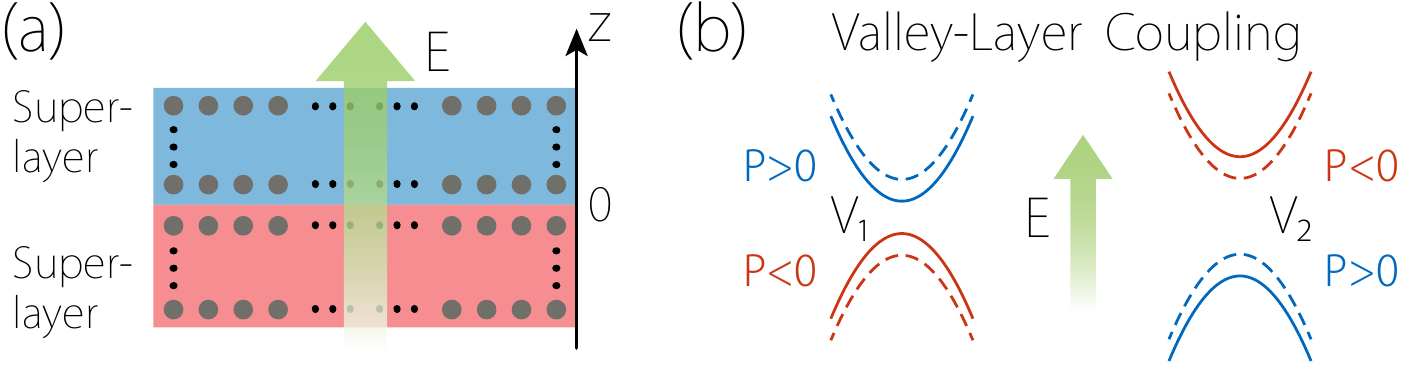}
\caption{
(a) A 2D material system viewed as consisting of two super-layers. Each super-layer may consist of a few atomic layers. (b) With valley-contrasting layer polarization ($P$), an applied gate electrical field can generate valley polarization.  The solid (dashed) curves denote the bands in the presence (absence) of the electric field. {$P>0$ ($P<0$) means the state is mainly localized on the top (bottom) super-layer.}
\label{fig1} }

\end{figure}

{\renewcommand{\arraystretch}{1.6}
\begin{table*}
\begin{ruledtabular}
\begin{tabular}{ccccc}
Lattice & BZ  & Valleys $\left(V_1,\ V_2\right)$ & Layer Group & Symmetry ${\cal O}$  \tabularnewline
\hline
\multirow{2}{*}{Centered  } & \multirow{5}{*}{\includegraphics[width=1.8cm]{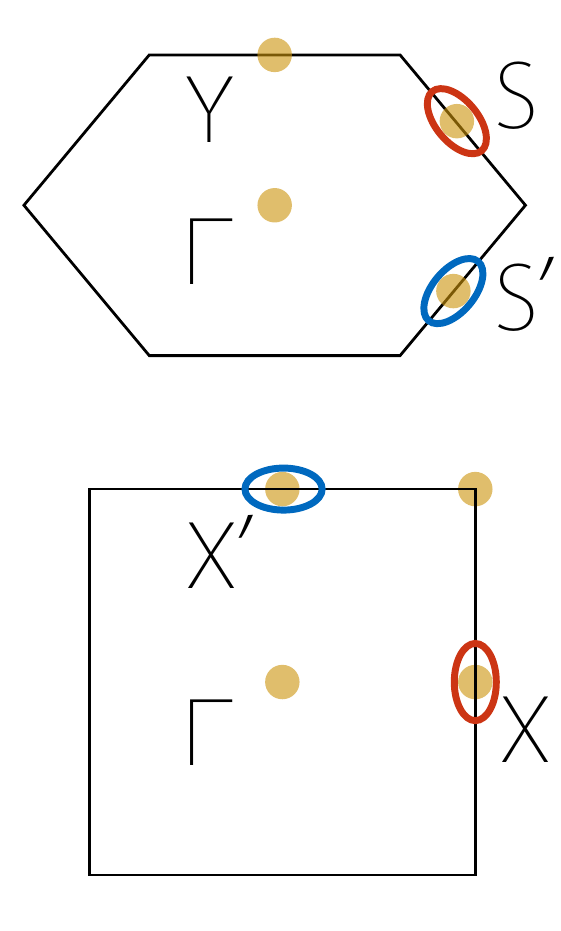}} & \multirow{2}{*}{$\left(S,\ S'\right)$} & 10 ($c211$) & $\left\{ C_{2x}|000\right\} $ \tabularnewline
Rectangular &  &  & 22 ($c222$) & $\left\{ C_{2x}|000\right\} $, $\left\{ C_{2y}|000\right\} $ \tabularnewline
\multirow{3}{*}{Square } &  & \multirow{3}{*}{$\left(X,\ X'\right)$} & 50 ($p\bar{4}$) & $\left\{ S_{4z}|000\right\} $ \tabularnewline
 &  &  & 59 ($p\bar{4}m2$) & $\left\{ S_{4z}|000\right\} $, $\left\{ C_{2,110}|000\right\} $ \tabularnewline
 &  &  & 60 ($p\bar{4}b2$) & $\left\{ S_{4z}|000\right\} $, $\left\{ C_{2,110}|\frac{1}{2}\frac{1}{2}0\right\} $ \tabularnewline
\end{tabular}
\end{ruledtabular}
\caption{\label{Tab1} Layer groups allowing for VLC. In the second column, the red and blue circles denote the two valleys $V_1$ and $V_2$. }
\end{table*}
}

Here, we present a completely new design principle for valleytronics, which goes beyond the existing paradigm to tackle this challenge. The idea is to explore systems in which the valleys are connected by a crystalline symmetry instead of $\mathcal{T}$, thereby circumventing the fundamental restriction posed by symmetry. Furthermore, we impose constraints on the crystalline symmetry, such that the system can admit a special coupling between valley and the layers that compose the system, which is termed as the valley-layer coupling (VLC). VLC enables a direct interaction between the valley and the gate electric field, as schematically illustrated in Fig. \ref{fig1} for generating valley polarization. We demonstrate our ideas via first-principles calculations and model analysis of an concrete 2D material example. Moreover, we show that systems with VLC exhibit additional interesting features, such as valley-contrasting linear dichroism and optical selection of (valley) interlayer excitons with a particular electric polarization.
Since the interlayer exciton has a dipole along the sample growth direction, it leads to the possibility of electric control and trapping of excitons, which can further facilitate the realization of exciton transistors or dipolar BEC~\cite{high2008,butov2002}.
Our work thus opens a completely new arena for valleytronics research.


\textit{\textcolor{blue}{Theory of valley-layer coupling.}} We consider a 2D material system. The system is extended in the $x$-$y$ plane, and has a finite thickness along $z$ (sometimes referred to as quasi-2D), which is the direction for the gate electric field. It can always be viewed as consisting of two super-layers: the top layer with $z>0$ and the bottom layer with $z<0$, where $z=0$ is set to be the midpoint of the system along $z$ [see Fig. \ref{fig1}(a)]. Note that each super-layer may actually contain multiple atomic layers.

In order to discuss VLC, we first define the super-layer polarization. For a Bloch band eigenstate $\psi_{n\bm k}(\bm r)$, we define its super-layer polarization $P_n(\bm k)$ as
\begin{equation}
P_n(\bm k)=\int_{z>0}|\psi_{n\bm k}|^2 d\bm r- \int_{z<0}|\psi_{n\bm k}|^2 d\bm r,
\end{equation}
where $n$ is the band index, $\bm k=(k_x,k_y)$ is wave vector in the 2D Brillouin zone (BZ), the position vector $\bm r$ has also a $z$ component due to the finite thickness, and one has $|P|\leq 1$ because of the normalization condition $\int |\psi_{n\bm k}|^2 d\bm r=1$. $P_n(\bm k)$ reflects the polarization of the state $\psi_{n\bm k}$ between the two super-layers: $P>0$ means the state has more weight distributed in the top layer; whereas $P<0$ means the distribution is biased towards the bottom layer. This quantity gives us a simple way to infer the behavior of the bands under a gate field.
As the vertical $E$ field produces a layer-dependent electrostatic potential, states with opposite $P$ are expected to acquire opposite energy shifts.

The concept of VLC can now be intuitively understood as the coupling between the valley degree of freedom and the super-layer polarization. Analogous to the valley-spin coupling in MoS$_2$-family materials \cite{XiaoPRL2012}, VLC requires the valleys to exhibit valley-contrasting super-layer polarizations. If realized, as we argued above, it will naturally achieve the electric control of valleys using gate field, as illustrated in Fig. \ref{fig1}(b).

Next, we analyze the symmetry requirements for the system to host VLC.
We consider nonmagnetic systems with a binary valley degree of freedom, which means that $\mathcal{T}$ is preserved and there are two inequivalent valleys in the BZ, denoted as $V_1$ and $V_2$. As we argued in the Introduction, to achieve valley polarization via gate field, the two valleys must not be connected by $\mathcal{T}$. This requires that each valley must be located at a time-reversal-invariant momentum (TRIM) point. (Recall that a 2D BZ has four TRIM points.) Moreover, The two valleys must be connected by certain crystalline symmetry operation $\mathcal{O}$ to enforce their degeneracy in energy. Namely,
\begin{equation}
\mathcal{O}:\qquad V_1\rightleftharpoons V_2.
\end{equation}
Then to enable VLC, the following considerations should be satisfied.

(i) The two valleys must have \emph{finite} and \emph{opposite} super-layer polarizations. This indicates that the operation $\mathcal{O}$ must satisfy the following relation:
\begin{equation}
  \mathcal{O} P(V_1) \mathcal{O}^{-1}=-P(V_2).
\end{equation}
Here, we drop the band index in $P$, which is understood to be either the lowest conduction band or the highest valence band. This requirement also implies that $\mathcal{O}$ must be broken by the applied gate electric field.

(ii) To have a finite $P$ at $V_1$ and $V_2$, certain crystalline symmetries must be broken. These include the horizontal mirror reflection $\mathcal{M}_z$ and the inversion $\mathcal{I}$. Because under their transformations
\begin{eqnarray}
\mathcal{M}_{z}P(V_i)\mathcal{M}_{z}^{-1}=-P(V_i), \\
\mathcal{I}P(V_i)\mathcal{I}^{-1}=-P(V_i),
\end{eqnarray}
where $i\in\{1,2\}$ and we have used the fact that $V_1$ and $V_2$ are at TRIM points (which are invariant under these operations), the polarization will be suppressed at the valley if either $\mathcal{M}_z$ or $\mathcal{I}$ is preserved.

Based on the above considerations, one can go through the 80 layer groups (LGs) for 2D materials to screen out the ones that allow the existence of VLC. The results are listed in Table~\ref{Tab1}. One can see that the candidates belong to two 2D Bravais lattices: the centered rectangular lattice (LG 10 and 22) and the square lattice (LG 50, 59, and 60). For LG 10 and 22, the two valleys should appear at TRIM points $S$ and $S'$, and they are connected by a twofold rotation with in-plane axis. For LG 50, 59, and 60, the two valleys should sit at $X$ and $X'$, and they are connected by a rotoreflection $\mathcal{S}_{4z}$ (for LG 59 and 60, a twofold rotation along $[110]$ also connects them). For a 2D material falling into one of these groups, as long as its conduction or valence band edge occurs at the requested TRIM points, it will generally host VLC and permit electric control of the valley degree of freedom.

\begin{figure}
\includegraphics[width=8.5cm]{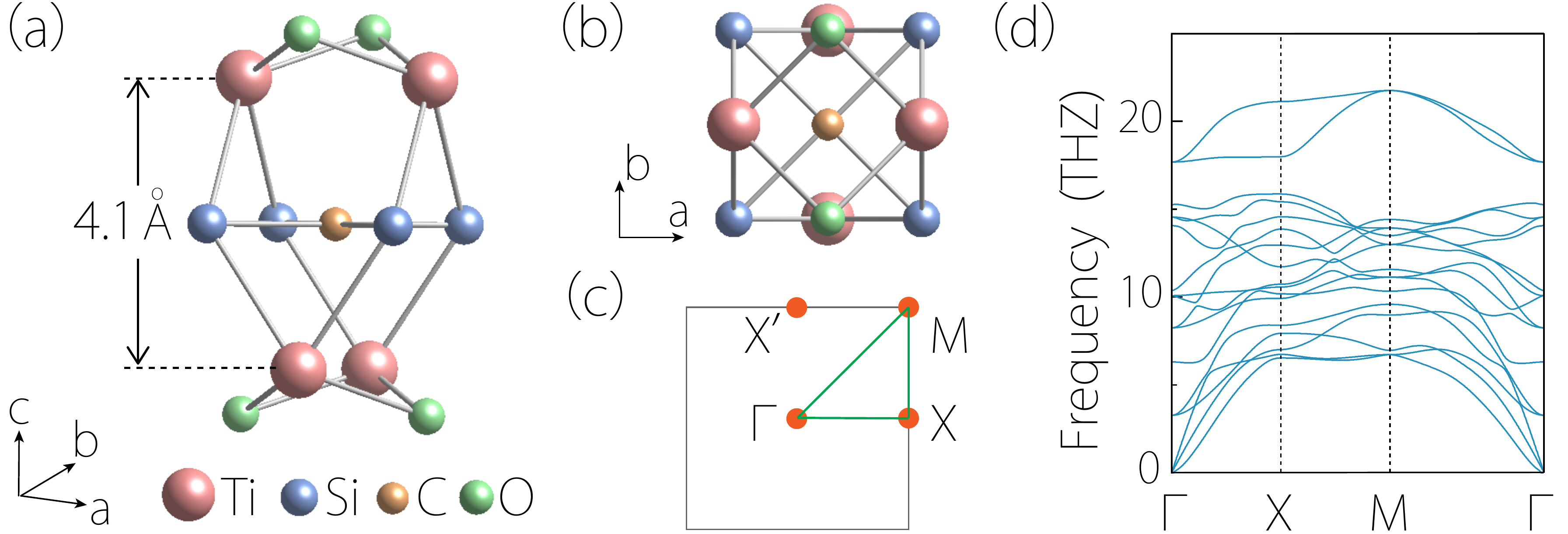}

\caption{(a) Side view and (b) top view of the crystal structure for monolayer TiSiCO. (c) The corresponding BZ. (d) Phonon spectrum for ML-TSCO. \label{fig2} }
\end{figure}

\textit{\textcolor{blue}{Application to a concrete example.}} Having
established the theory of VLC, below, we illustrate our key idea with a concrete material example.
The symmetry conditions obtained above (Table~\ref{Tab1}) offers useful guidance in searching for suitable material systems.
Here, we consider the monolayer TiSiCO (ML-TSCO). As illustrated in Fig.~\ref{fig2}, it has similar structure as ML-HfGeTe~\cite{GuanPRM2017} and ML-ZrSiO studied before~\cite{XuPRB2015}.
It possesses five square-lattice atomic layers vertically stacked in the sequence of O-Ti-Si/C-Ti-O, with LG No.~59 ($P\bar{4}m2$), which is one candidate LG in Table~\ref{Tab1}.
Via first-principles calculations, we confirm that the monolayer is dynamically stable [see Fig.~\ref{fig2}(d)] and also enjoys good thermal stability up to 800 K~\cite{SM}.
The optimized lattice constant is $a=b=2.817$\AA. As we shall see, the low-energy states in ML-TSCO is mainly distributed in the two Ti atomic layers. It is worth noting that the two Ti layers have a separation about
4.1 \AA, which is even comparable to the typical interlayer separation for van der Waals stacked heterostructures.

\begin{figure}
\includegraphics[width=8.5cm]{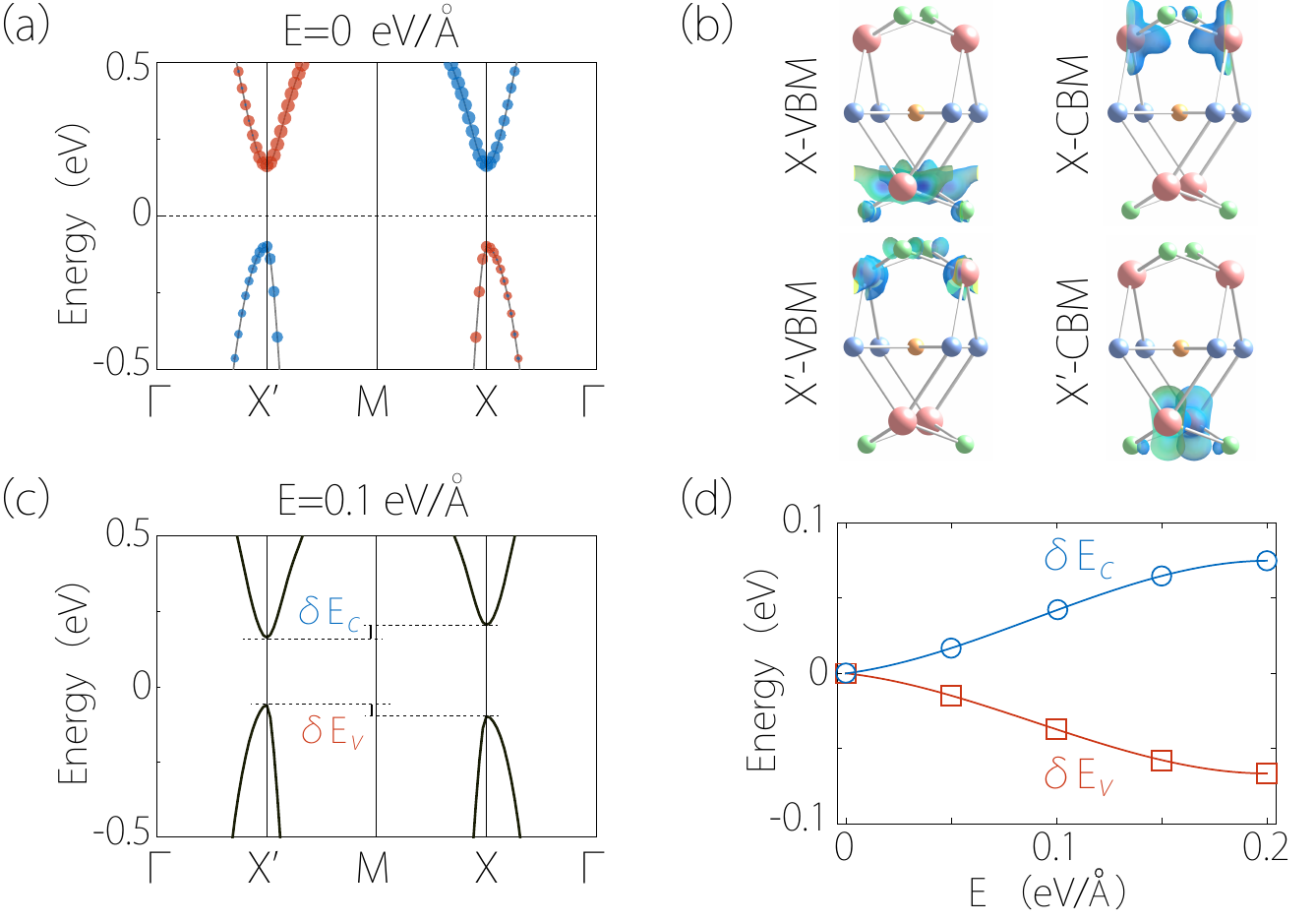}

\caption{(a) Electronic band structure for ML-TSCO. The size of the red (blue) dots in (a) is proportional to the weight of projection onto atomic orbitals in the bottom (top) Ti atoms. (b) Charge density distribution plotted for VBM and CBM states at the two valleys. (c) shows the band structure under a gate field of $E=0.1$ eV/{\AA}.
(d) Valley splitting for VBM ($\delta E_{v}$) and CBM ($\delta E_{c}$) [indicated in (c)] versus the applied gate field.  \label{fig3} }
\end{figure}

According to Table~\ref{Tab1}, a material with LG~59 can host VLC if it has valleys located at $X$ and $X'$ points. This is indeed the case for ML-TSCO. Figure~\ref{fig3}(a) shows the calculated electronic band structure. Here, the spin-orbit coupling (SOC) is not included because of its negligible strength for this material. One clearly observes that a pair of valleys for both conduction and valence bands occur at $X$ and $X'$.
The two valleys are connected by $\mathcal{S}_{4z}$ and $\mathcal{C}_{2,[110]}$, rather than $\mathcal{T}$. Hence, according to our previous analysis, they must have finite and opposite super-layer polarizations. This is confirmed by the first-principles calculation. By analyzing the spatial distribution of the states [see Fig. \ref{fig3}(a-b)], one finds the following features. (i) The band edge states are mainly distributed in the two Ti layers.
(ii) The conduction band at $X$ valley is mainly distributed in the top layer, whereas for the $X'$ valley it is mainly in the bottom layer. (iii) The layer distribution pattern for the valence bands is reversed. The physical picture is similar to the schematic illustration in Fig. \ref{fig1}(b). Thus, the first-principles calculation confirms the existence of VLC, consistent with our theory.

To demonstrate that VLC indeed enables the gate field control of valleys, we explicitly calculate the band structure under a vertical electric field.
As shown in Fig. \ref{fig3}(c), for the conduction band, the $E$ field pulls down the $X'$ valley and pushes up the $X$ valley, successfully generating a valley polarization for the electron carriers when the system is $n$-doped. For the valence band, the valley shifts are opposite, due to the reversed super-layer polarization, so valley polarization can also be generated for holes when the system is $p$-doped. Our calculation shows that for ML-TSCO, a gate field of 0.1 eV/\AA\ (achievable with current experimental technique~\cite{zhang2009}) can induce a valley energy splitting of $37$ meV and $42$ meV for valence band and conduction band, respectively [see Fig.~\ref{fig3}(d)]. Such energy scale is much larger than the thermal energy at room temperature (25 meV), suggesting its potential for room temperature device operations.

\textit{\textcolor{blue}{Effective model for VLC.}} To better understand the physics of VLC, we develop an effective model based on the above result for ML-TSCO. We stress that the essential features of the model are general, not depending on the particular material, because they are determined only by symmetry. Here, the low-energy physics occur at $X$ and $X'$ points, both having the little group symmetry of $C_{2v}$.  At $X$
($X'$) point, the two low-energy bands correspond to the 1D representations $A_{1}$ and $B_{2}$ ($A_{1}$ and $B_{1}$) for the
$C_{2v}$ group. Using them as
basis states, the $k\cdot p$ effective model expanded up to linear order reads~\cite{SM}
\begin{eqnarray}
{\cal H}_{0} & = & \Delta\sigma_{z}+v(k_{x}+k_{y})\sigma_{y}-v(k_{x}-k_{y})\tau_{z}\sigma_{y},\label{eq:ham}
\end{eqnarray}
where the wave vector $\bm k$ is measured from each valley center, $\Delta$ represents the band gap, $\tau_{z}=\pm1$
denotes the $X$/$X'$ valley, and $\sigma$'s are Pauli matrices acting on the basis states. Since the two basis are respectively polarized
in the top and bottom layers, $\sigma$ can also be regarded as acting on the layer index space, and in particular, the super-layer polarization for a state $\psi$ is given by $P_\psi=\langle\psi|\sigma_z|\psi\rangle$. The last term in Eq.~(\ref{eq:ham}) involving $\tau_{z}\sigma_{y}$ clearly indicates a coupling between
valley and layer degrees of freedom.
The effect of vertical electric field then can be captured
by
\begin{eqnarray}
{\cal H}_{E} & = & \Delta_{E}\tau_z\sigma_{z},\label{eq:hamE}
\end{eqnarray}
where the effective strength $\Delta_{E}$ may be taken as linear in the $E$ field, \emph{i.e.}, $\Delta_{E}=\alpha E$, in the simplest approximation.
It follows that the applied $E$ field shifts the band edges at the two valleys in opposite ways, and the local gaps at the two valleys become $(\Delta\pm \Delta_E)$. The model parameters $\Delta$, $v$, and $\alpha$ can be obtained by fitting the DFT results. The obtained values for ML-TSCO are $\Delta=0.133$ eV, $c_3=0.59$ eV$\cdot$\AA ~and $\alpha \approx0.2$ \AA.

\begin{figure}
\includegraphics[width=8.5cm]{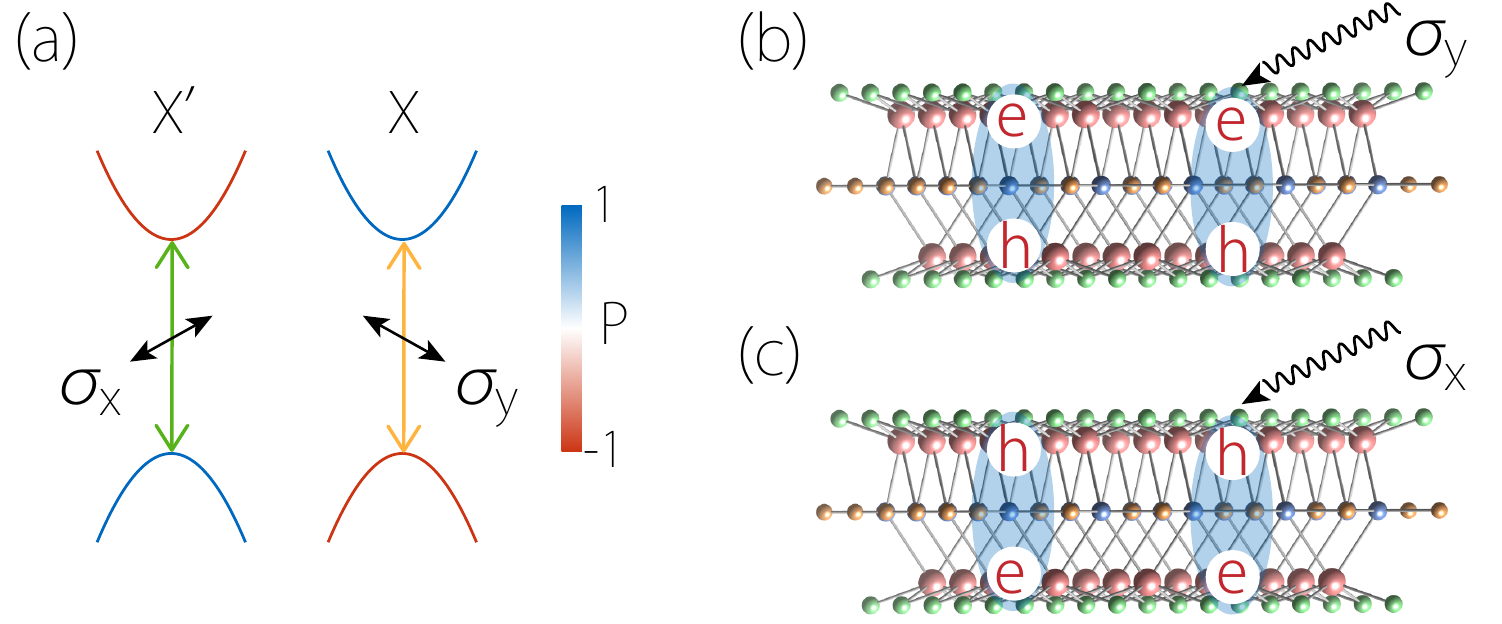}

\caption{
(a) Valley optical transition selection rules. (b) $X$ and (c) $X'$ valley-polarized interlayer excitons with opposite electric polarization can be selectively excited by $\sigma_y$ and $\sigma_x$ linear polarized optical fields, respectively. \label{fig4} }
\end{figure}

\textit{\textcolor{blue}{Valley linear dichroism and interlayer exciton.}} Since the two valleys are characterized by $C_{2v}$ symmetry and are connected by $\mathcal{S}_{4z}$, they should exhibit valley-contrasting linear dichroism in the optical interband absorption.
This can be readily verified by using the effective model in Eq.~(\ref{eq:ham}). The coupling strength with optical fields linearly polarized in the $i$-th direction is given by
\begin{equation}
M_{i}(\bm{k})=m_e\langle u_{c}(\bm{k})|\frac{\partial{\cal H}}{\partial k_{i}}|u_{v}(\bm{k})\rangle,
\end{equation}
where $m_e$ is the free electron mass, and $u_{c(v)}$ the periodic part of the Bloch state of conduction
(valence) band. With the model in Eq.~(\ref{eq:ham}), we find that for the $X$ valley
\begin{equation}
|M_x^X(\bm k)|^2=O(k^2),\qquad |M_y^X(\bm k)|^2=\frac{4v^{2}m_e^2\Delta^{2}}{4v^{2}k_{x}^{2}+\Delta^{2}}.
\end{equation}
Similarly, for the $X'$ valley, we have
\begin{equation}
  M_{x/y}^{X'}(k_{x},k_{y})=M_{y/x}^{X}(k_{y},-k_{x}).
\end{equation}
Close to the valley center, the interband transitions are coupled exclusively with $x$-linearly
($y$-linearly) polarized light for the $X'$ ($X$) valley, consistent with our symmetry argument, as shown in Fig.~\ref{fig4}(a). This means that we can selectively excite carriers in one valley by controlling the polarization of light.
Applying gate electric field will not affect the linear dichroism, but can further split the band-edge transition frequency by $\pm \Delta_E$ for the two valleys.

Here, due to the super-layer polarization of the valley states, for a particular valley, the optically excited electrons and holes are located at different layers. The interaction between the electrons and holes leads to the formation of excitons. Interestingly, the excitons here are interlayer excitons with valley-contrasting layer polarization. As illustrated in Fig.~\ref{fig4}(b), the $x$-polarized light with the band-edge excitonic transition frequency  will selectively generate excitons in the $X'$ valley, with holes in the top layer and electrons in the bottom layer. The situation is reversed for the $y$-polarized light [see Fig.~\ref{fig4}(c)]. Note that these interlayer excitons naturally carry a charge polarization along $z$, so they can be electrically manipulated \cite{jones2014,rivera2015,rivera2016,wang2017,ross2017} and the effect of linear dichroism can also be probed by voltage measurement across the thickness.

\emph{\textcolor{blue}{Discussion.}} In this work, we have proposed a new design principle to achieve static control of valley polarization by gate electric field.
The essence is to switch the focus from conventional $\mathcal{T}$-connected valleys to valleys connected by crystalline symmetry, and to request the system to have the right symmetry to allow a new coupling mechanism---the VLC. We point out that such electrically generated valley polarization may also permit a purely electrical detection. Note that each valley in LG 59 and 60 only has twofold rotational symmetry, indicating that the dispersion in the valley must be anisotropic. For example, in ML-TSCO, for the $X$ valley, the effective mass of valence band along $x$ is $m_x^*\approx0.84m_e$, whereas $m_y^*\approx0.16m_e$, differing by $\sim 5$ times. Due to the $\mathcal{S}_{4z}$ symmetry, the anisotropy at the $X'$ valley is rotated by $\pi/2$, so that in the absence of valley polarization, the transport properties are still isotropic. The valley polarization breaks the $\mathcal{S}_{4z}$ symmetry, making the transport anisotropic. For example, ML-TSCO with $X$ valley polarization should have a larger resistance along $x$ than $y$. By measuring the anisotropic resistance, one can detect the valley polarization. Thus, both valley information write-in and read-off can be achieved by purely electric means, which is a great advantage of our scheme.

Another advantage of our proposal is that the valleys in the target systems are located at TRIM points, which are well separated in $k$ space. For example, for LG 59 and 60, the separation between $X$ and $X'$ represents the longest distance between any two points in the BZ. This is beneficial for increasing the carriers' inter-valley scattering time, making valley index a good information carrier for valleytronics applications.

The optical properties revealed above would endow the system with additional possibilities. For example, the valley polarization can also be detected by the difference in the optical absorbances for $x$ and $y$ polarized lights. In addition, the interlayer excitons, which can be selectively created here by linearly polarized light, are expected to have longer lifetime due to the spatial separation between electron and hole.  The long lifetime is desirable for exciton valleytronics~\cite{Schaibley2016,YuPRL2015,onga2017} and for achieving more exotic phases such as exciton condensation~\cite{eisenstein2004,high2012}.

\bibliographystyle{apsrev4-1}
\bibliography{VLC_ref}

\end{document}


\title{Supplemental Materials for ``Valley-Layer Coupling: A New Design Principle for Valleytronics"}

\author{Zhi-Ming Yu}
\address{Research Laboratory for Quantum Materials, Singapore University of Technology and Design, Singapore 487372, Singapore}

\author{Shan Guan}
\address{Research Laboratory for Quantum Materials, Singapore University of Technology and Design, Singapore 487372, Singapore}
\affiliation{State Key Laboratory of Superlattices and Microstructures, Institute of Semiconductors, Chinese Academy of Sciences, Beijing 100083, China}

\author{Xian-Lei Sheng}
\address{Department of Physics, Key Laboratory of Micro-nano Measurement-Manipulation and Physics (Ministry of Education), Beihang University, Beijing 100191, China}
\address{Research Laboratory for Quantum Materials, Singapore University of Technology and Design, Singapore 487372, Singapore}

\author{Weibo Gao}
\address{Division of Physics and Applied Physics, School of Physical and Mathematical Sciences, Nanyang Technological University, Singapore 637371, Singapore}
\address{The Photonics Institute and Centre for Disruptive Photonic Technologies, Nanyang Technological University, Singapore 637371, Singapore}

\author{Shengyuan A. Yang}
\address{Research Laboratory for Quantum Materials, Singapore University of Technology and Design, Singapore 487372, Singapore}

\maketitle

\section{Details of First-Principles Calculation}

Our first-principles calculations were based on the density functional theory (DFT), carried out using the projector augmented wave (PAW) method \cite{Blochl1994} as implemented in the Vienna ab-initio simulation package (VASP) \cite{Kresse1993,Kresse1996}. The generalized gradient approximation with the Perdew-Burke-Ernzerhof (PBE) \cite{Perdew1996} realization was adopted for the exchange-correlation functional. The obtained band structures were calculated by the  hybrid functional approach (HSE06)~\cite{Heyd2003}. The cutoff for plane-wave expansion was set to be 520 eV. The Monkhorst-Pack $k$-point mesh with a size of $20\times20\times1$ was used to sample the Brillouin zone. And a vacuum layer of 18 {\AA} thickness was added to avoid artificial interactions between periodic images. The convergence criteria for energy and force were set to be 10$^{-6}$ eV and 0.005 eV/{\AA}, respectively.

\begin{figure}[h]
\includegraphics[width=6cm]{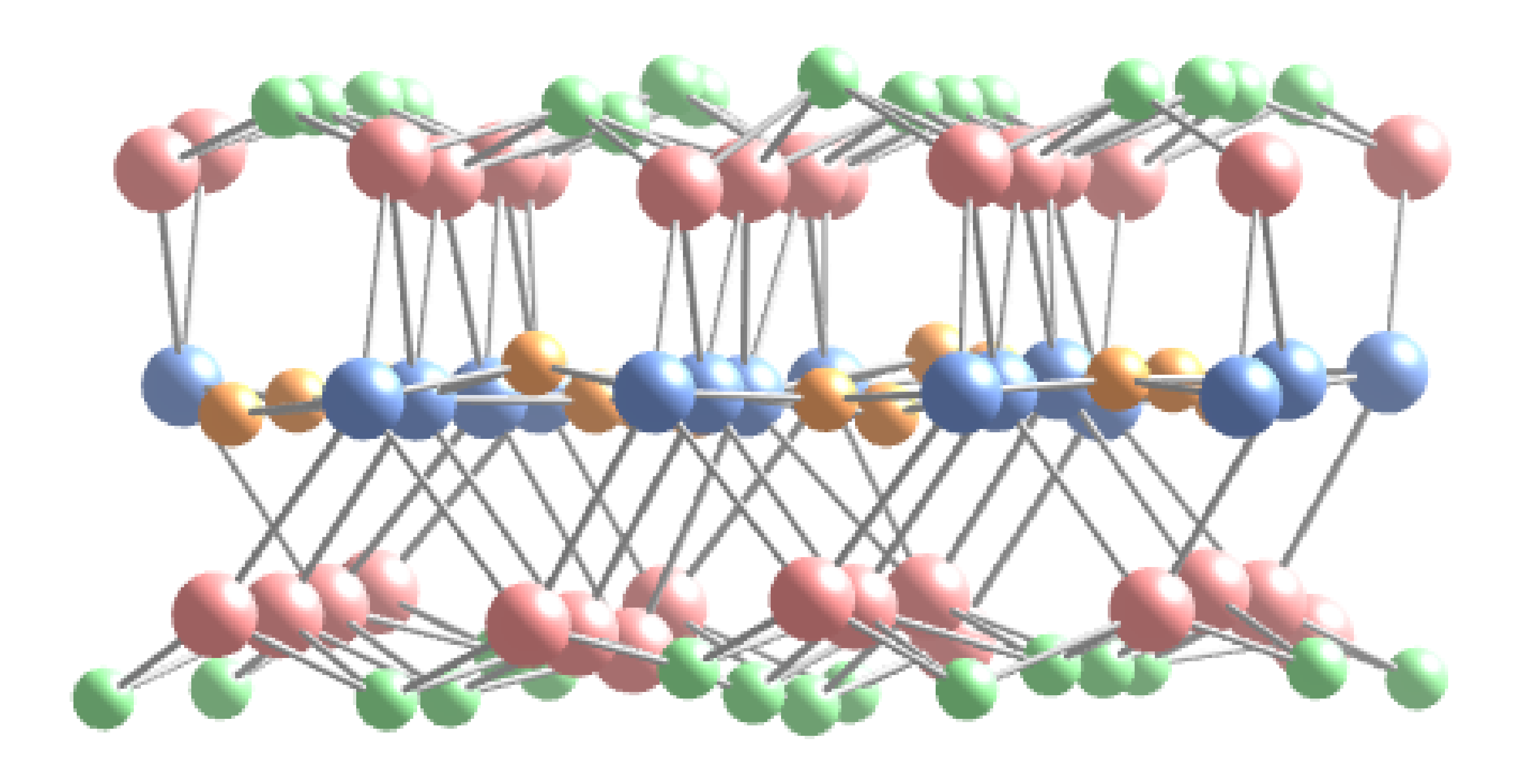}
\caption{The snapshot taken at the last step (2 ps) in the time evolution of first-principles molecular dynamics simulation at 800 K, showing that the structure is maintained.
 \label{figs1} }
\end{figure}

To confirm the dynamical stability of ML-TSCO, the phonon spectrum was calculated by the Phonopy \cite{Togo2008} code with a supercell of $6\times6$ unit cells. To check the thermal stability, the first-principles molecular dynamics simulations within canonical ensemble at finite temperature were performed on a $4\times4$ supercell with time duration up to 2 ps and a step interval of 0.5 fs. The results indicate that the lattice structure of ML-TSCO remains stable at 800 K, as shown in Fig. \ref{figs1}.

\section{Derivation of effective model }

We present the derivation of the low-energy effective model shown in Eq.~(6) in the main text. The little group at $X$ ($X'$) is $C_{2v}$. The effective Hamiltonian for the valley electrons is constrained by the generators of $C_{2v}$, namely, the two mirrors ${\cal M}_{x}$ and ${\cal M}_{y}$, and also by the time reversal symmetry ${\cal T}$. The CBM and VBM states at $X$ have the symmetries of $A_{1}$ and $B_{1}$ representations for $C_{2v}$. With them as basis,
the symmetry operators have the following representations:
\begin{eqnarray}
{\cal M}_{x}=\sigma_{z},\ \  & {\cal M}_{y}=\sigma_{0},\ \  & {\cal T}={\cal {K}},
\end{eqnarray}
with ${\cal {K}}$ the complex conjugation operator and $\sigma_{0}$ the $2\times2$ identity matrix. The effective Hamiltonian must satisfy the following relations
\begin{eqnarray}
{\cal M}_{x}^{-1}{\cal H}^{X}(k_x,k_y){\cal M}_{x} & = & {\cal H}^{X}(-k_{x},k_{y}),\\
{\cal M}_{y}^{-1}{\cal H}^{X}(k_x,k_y){\cal M}_{y} & = & {\cal H}^{X}(k_{x},-k_{y}),\\
{\cal T}^{-1}{\cal H}^{X}(\bm k){\cal T} & = & {\cal H}^{X*}(-\bm k).
\end{eqnarray}
Here, $\bm k$ is measured from the $X$ point. Thus, expanded to second order in $k$, we obtain
\begin{eqnarray}\label{hamX}
{\cal H}^{X} & = & W^X({\bm k})+(\Delta +c_1 k_x^2+c_2 k_y^2)\sigma_z+c_3 k_x\sigma_y ,
\end{eqnarray}
where $W^X({\bm k})=w_0+w_{x}k_{x}^{2}+w_{y}k_{y}^{2} $. Since $X$ and $X'$ valley are connected by $S_{4z}$, the effective Hamiltonian for $X'$ valley can be obtained as
\begin{eqnarray}\label{hamY}
{\cal H}^{X'}(k_x,k_y) & = & {\cal H}^{X}(k_y,-k_x).
\end{eqnarray}
The expressions in (\ref{hamX}) and (\ref{hamY}) can be combined into a single equation:
\begin{eqnarray}
{\cal H}_{0} & = & w_0+\alpha_{+}k^{2}+\left(\Delta+\beta_{+}k^{2}\right)\sigma_{z}+v(k_{x}+k_{y})\sigma_{y}
+\left[(\alpha_{-}+\beta_{-}\sigma_{z})(k_{x}^{2}-k_{y}^{2})+v(k_{x}-k_{y})\sigma_{y}\right]\tau_{z},\label{eq:ham}
\end{eqnarray}
where $k=\sqrt{k_x^2+k_y^2}$, $\alpha_{\pm}=(w_x\pm w_y)/2$, $\beta_{\pm}=(c_1\pm c_2)/2$ and $v=c_3/2$.
Keeping terms up to $k$-linear order in Eq.~(\ref{eq:ham}) recovers the effective Hamiltonian in Eq.~(6) in the main text. The model parameters can be obtained by fitting the DFT results. Specifically, for ML-TSCO, we obtain $w_0=0.0328$ eV, $\Delta=0.133$ eV, $w_x=0.714$ eV$\cdot$\AA$^2$, $w_y=-9.046$ eV$\cdot$\AA$^2$, $c_1=0$, $c_2=14.522$ eV$\cdot$\AA$^2$ and $c_3=1.183$ eV$\cdot$\AA.